# Hidden Density-Wave Instability in the Trimer Ruthenate $Ba_4Ru_3O_{10}$

Gang Cao[1*], Hengdi Zhao[1,2], Adrienne Bond[1], Tristan R. Cao[1], Gabriel Schebel[1], Arabella Quane[1], Yifei Ni[1], Yu Zhang[1], Logan Wall[1], Rahul Nandkishore[1], Pedro Schlottmann[3], Stephan Rosenkranz[2], and Feng Ye[4]

[1]Department of Physics, University of Colorado at Boulder, Boulder, CO 80309, USA

[2]Materials Science Division, Argonne National Laboratory, Lemont, IL 60439, USA

[3]Department of Physics, Florida State University, Tallahassee, FL 32306, USA

[4]Oak Ridge National Laboratory, Oak Ridge, TN 37830, USA

**Abstract**

We report an unprecedented hidden density-wave instability in the trimer-based ruthenate $Ba_4Ru_3O_{10}$, previously regarded as a purely antiferromagnetic insulator. This instability develops in two distinct stages: an electronically driven reconstruction at $T_A$ = 100 K manifested in structural, thermodynamic, and transport anomalies that remain remarkably insensitive to magnetic fields up to at least 14 T, followed only at much lower temperatures T* ~ 20 K by the emergence of strongly nonlinear transport. Below T*, charge conduction exhibits distinct depinning thresholds, sharp negative differential resistance, and unusually slow collective dynamics in the Hertz range. Direct measurements show that Joule heating is negligible, and all nonlinear signatures vanish upon only 3% Ir substitution for Ru, demonstrating the intrinsic origin. These results identify $Ba_4Ru_3O_{10}$ as a rare correlated system hosting a strongly pinned collective electronic mode intertwined with antiferromagnetism.

*gang.cao@colorado.edu

Collective charge density waves (CDWs) are a fundamental manifestation of electronic self-organization, typically realized in low-dimensional weakly correlated materials [1-17]. In contrast, strongly correlated transition-metal oxides more often favor magnetism or Mott localization, making clear examples of sliding collective charge transport exceedingly rare [18-20]. Identifying density-wave-like instabilities in correlated oxides is therefore of particular interest, as they may represent unconventional electronic reconstructions intertwined with magnetism.

In this Letter, we uncover a density-wave instability in the trimer ruthenate $Ba_4Ru_3O_{10}$, previously regarded as a purely antiferromagnetic (AFM) insulator. Unlike a conventional CDW, this instability is thermodynamically established at $T_A$ = 100 K but dynamically revealed only at T* ~ 20 K (**Fig. 1a**). The transition at $T_A$ produces clear structural, thermodynamic, and transport anomalies yet remains insensitive to magnetic fields H up to at least 14 T, pointing to an electronically driven reconstruction. Remarkably, only at T* ~ 20 K, transport becomes strongly nonlinear, exhibiting distinct depinning thresholds, negative differential resistance (NDR), and slow collective dynamics, consistent with a strongly pinned collective mode. The inferred relaxation time is on the order of $10^{-3}$ s, many orders of magnitude longer than $10^{-15}$- $10^{-12}$ s typical of conventional CDWs [1], highlighting the unusually strong pinning of the collective electronic state. Direct detection of an ordering wavevector remains an important direction for future scattering experiments.

Although depinning is a generic feature of CDWs and NDR has been reported in a limited number of cases [21-30], the simultaneous observation of distinct depinning and NDR thresholds together with unusually slow collective dynamics in a magnetic oxide is unprecedented. Notably, all nonlinear transport signatures vanish upon only 3% Ir substitution for Ru, which preserves the crystal structure and magnetic insulating state, strongly arguing against ruling out Joule heating or extrinsic artifacts and demonstrating the intrinsic fragility of the density-wave instability. See direct measurements of current-induced temperature changes in Supplemental Material (SM).

Prior studies of $Ba_4Ru_3O_{10}$ have focused primarily on its magnetic properties [31-33]. μSR and neutron scattering established AFM order below ~ 105 K with a low-spin S = 1 state [33]. However, the field independence of $T_A$ and the emergence of pronounced nonlinear transport below T* cannot be explained within a purely magnetic framework, pointing instead to a density-wave instability developing deep inside the AFM phase. The wide separation between $T_A$ and T* naturally explains why this density-wave instability has escaped detection in earlier magnetic probes (**Fig. 1a**).

*Crystal structure and alterations near transition.* $Ba_4Ru_3O_{10}$ crystallizes in an orthorhombic *Cmca* structure composed of $Ru_3O_{12}$ trimers of face-sharing $RuO_6$ octahedra (**Fig. 1b-1d**), consistent with previous work [31]. These trimers form corrugated two-dimensional sheets in the *ac* plane that are stacked along the *b* axis. Clear anomalies appear near $T_A$ in the unit-cell volume V, Ru–Ru distances, and Ru–O–Ru bond angles (**Fig. 1e-1g,** SFig.1 in SM), indicating lattice involvement in the electronically driven transition [33].

Within each trimer, the middle Ru1 and outer Ru2 sites are inequivalent: Ru1 exhibits longer Ru-O bonds, implying substantial charge disproportionation between the two sites [32]. Such molecular-like trimer units provide additional internal degrees of freedom and offer a natural platform for unconventional electronic instabilities in high-Z correlated oxides (Z being atomic number) [18, 19, 32, 37-41].

*Magnetic properties, heat capacity and their insensitivity to magnetic fields.* The data in **Fig. 2** establish that $T_A$ = 100 K in $Ba_4Ru_3O_{10}$ is largely driven by an intrinsic electronic instability.

Above $T_A$, $\chi(T)$ for all three directions exhibits no Curie-Weiss behavior but weak temperature dependence up to at least 350 K, indicating the dominance of Pauli-like moments associated with itinerant electrons rather than localized magnetic moments (**Fig.2a**). $\chi(T)$ becomes more anisotropic below $T_A$, with the easy axis along the *a* axis (**Fig.2a**). Importantly, $T_A$ is insensitive to H up to at least 14 T, as demonstrated by both $\chi(T)$ (**Fig. 2b**) and heat capacity C(T) (**Fig. 2d**). In addition, the isothermal magnetization M(H) further supports an electronically driven instability. As shown in **Fig. 2c**, the slope of *a*-axis $M_a(H)$ at 1.8 K is substantially reduced compared to that at 150 K; for example, at 7 T, $M_a$ at 150 K is nearly twice that at 1.8 K. Such behavior is inconsistent with an insulating local-moment magnet, for which M(H) ~ 1/T, but is instead consistent with reduction of low energy density of states associated with an electronic reconfiguration below $T_A$, which weakens the Pauli susceptibility and M below $T_A$. (In contrast, $M_a$ for 3% Ir doped $Ba_4Ru_3O_{10}$ recovers the normal behavior M(H) ~ 1/T, see SFig. 2 in SM.) Consistently, the transition at $T_A$ results in only a small entropy removal of $\Delta S \approx 0.4$ J/mole K, which is only ~ 4% of the full spin entropy Rln (2S+1) = Rln 3 per Ru (R = universal gas constant). Interestingly, the value of $\Delta S$ is identical to that in the architype CDW $K_{0.3}MoO_3$ [42]. In addition, **Fig. 2f** shows the absence of a $T^3$-term anticipated for an AFM ground state in C(T) below 10 K. These behaviors reinforce an electronic nature of the transition at $T_A$.

*Nonlinear transport and effects of light Ir doping.* The transport behavior reveals a crossover from Ohmic to nonlinear transport (shaded area in **Fig. 3c**). At T > 20 K, the *a*-axis resistivity $\rho_a(T)$ is weakly current-dependent. However, below T* ~ 20 K, $\rho_a(T)$ becomes extraordinarily sensitive to the applied current I (**Fig.3c**), e.g., increasing I from 0.1 mA to 5 mA leads to a reduction of $\rho_a$ by nearly 3 orders of magnitude below T* (**Fig. 3e,** blue curve). Consistently, the I-V curve (blue) exhibits a distinct depinning threshold $V_{DP}$ and a sharp $V_{NDR}$ (**Fig. 3f)**. All these signal the emergence of a collective conduction mode that dominates transport at T* < 20 K. The delay of the occurrence of the nonlinear transport is likely because above T*, transport is dominated by thermally activated quasiparticles, which provide an efficient parallel conduction channel and obscure the collective response of the density-wave instability. As temperature is lowered, these quasiparticle channels progressively freeze out, leading to enhanced nonlinearity governed by the strongly pinned collective mode (**Fig.1a**). Relevantly, the resistivities along the *a* and *c* axes merge below T* despite the strongly anisotropic crystal structure (**Inset** in **Fig.3c**) (Note that ρ(T)/ρ(300K) is presented for a more accurate comparison between $\rho_a$(T) and $\rho_c$(T).), reinforcing a crossover from a quasiparticle-dominant regime (>T* ~ 20 K) into the collective transport regime. In this regime, charge transport is no longer governed by band velocities or single-particle scattering rates, which dictate Ohmic transport above T*, but by pinning, depinning dynamics or the coordinated motion of the instability (**Fig.1a**).

It is important to emphasize that no anomaly is observed near T* in the heat capacity (**Fig. 2e**), magnetic susceptibility (**Figs. 2a,b**), or resistivity (**Fig. 3b**), indicating that T* does not correspond to a phase transition but rather marks a crossover into a collective transport regime. This conclusion is further supported by our additional x-ray data down to 20 K (Table S1-S3 in SM) and a prior structural study, which reports no additional lattice anomaly down to 10 K [31]**.**

Furthermore, a thermal activation gap Δ ~ 34 meV is inferred from $\rho_a$(T) at I = 0.1 mA below 100 K (**Inset of Fig.3c**), yielding a gap ratio $2\Delta/k_BT$ ~ 8 ($k_B$ = Boltzmann constant), far exceeding the weak-coupling CDW limit 3.5 - 4 [1], placing the electronic instability outside a conventional Peierls limit. Note that Δ ~ 34 meV represents a lower bound due to parallel conduction channels.

We introduce 3% Ir doped $Ba_4Ru_3O_{10}$ or $Ba_4(Ru_{0.97}Ir_{0.03})_3O_{10}$ in order to provide a crucial control for identifying the origin of the nonlinear transport and examining any role of Joule heating. **Figs. 3a-3b** compare the temperature dependence of $\chi_a$ and $\rho_a$ of pristine $Ba_4Ru_3O_{10}$ (x = 0) and $Ba_4(Ru_{0.97}Ir_{0.03})_3O_{10}$ (x = 0.03). Light Ir doping preserves the crystal and magnetic structures [40] and overall insulating behavior with a lower $T_A$ at 85 K (**Figs. 3a-3b,** SFig.2 in SM). However, the Ir-doped sample exhibits nearly Ohmic transport over the entire temperature range (**Fig. 3d**), with no sign of nonlinear I-V curves (**Fig.3f**, red I-V curve for x = 0.03). The suppression of nonlinear transport by minimal Ir substitution indicates that the underlying density-wave instability is a fragile electronic state highly susceptible to even slight lattice or electronic alterations (indeed, its sister trimer lattice $BaRuO_3$ is a good metal showing strong quantum oscillations [43]).

Remarkably, the absence of nonlinear transport in insulating $Ba_4(Ru_{0.97}Ir_{0.03})_3O_{10}$ (**Fig. 3b**) indicates that Joule heating plays no role in the nonlinear behavior observed in pristine $Ba_4Ru_3O_{10}$. Furthermore, our direct measurements of current-induced temperature changes, $\Delta T$, show that the heating is minimal: $\Delta T < 3.1$ K at 10 K even at I = 10 mA, and negligible (< 1 K) at I = 2 mA (SFig.3 in SM), where pronounced nonlinear transport is already present (**Fig. 4a**). Such small temperature variations, along with the smoothly varying resistivity $\rho(T)$ below 30 K (**Fig.3b**), cannot account for the highly anisotropic and sharply nonlinear I-V characteristics, which are therefore inconsistent with a Joule-heating origin (SM for more discussion). All these firmly establish the intrinsic electronic origin of the nonlinear transport.

*Anisotropy, temperature- and frequency-dependence of I-V characteristics of* $Ba_4Ru_3O_{10}$ *(x = 0).* We now examine I-V characteristics at T < T*. **Fig. 4a** presents the I-V characteristics for both the *a* and *c* axes at 10 K and 24.2 Hz. The I-V curve for the *a* axis (blue curve) features two distinct anomalies: a $V_{DP}$ and a $V_{NDR}$. $V_{DP}$ marks the onset of collective motion of a pinned electronic state, while $V_{NDR}$ signals its reorganization into a more conductive collective regime.

The I-V curves in **Fig. 4a** also reveal a pronounced anisotropy. While the *a*-axis I-V characteristics display both $V_{DP}$ and $V_{NDR}$, the *c*-axis response exhibits only $V_{NDR}$, with no well-defined $V_{DP}$. To eliminate geometric and sample-size effects, these data are replotted as current-density J - electric-field E curves in **Fig. 4b**. The large and persistent anisotropy in the characteristic fields $E_{NDR}$ between the two directions confirms that the differences observed in the I-V curves are intrinsic. This behavior indicates that the instability is strongly direction dependent, with the *c* axis providing a much lower-threshold channel for the onset of collective motion; e.g., $E_{NDR}$ = 13 V/cm, $V_{NDR}$ = 0.5 V at 10 K and 24.2 Hz for the *c* axis, compared to $E_{NDR}$ = 58 V/cm, $V_{NDR}$ = 3 V for the *a* axis, indicating that the *a*-axis transport requires substantially larger driving fields to enter into a reorganized NDR regime (**Figs. 4a-4b**). The much lower values of $E_{NDR}$ and $V_{NDR}$ for the *c* axis are likely associated with a more favorable connectivity along that direction with zigzag chains (**Fig.1c**). Note the *a*-axis $E_{DP}$ (~ 30 V/cm in **Fig. 4b**) is much larger than those in other CDWs, such as $TaS_3$ whose $E_{DP}$ ~ 0.3 V/cm [1], confirming the strong pinning in the collective mode. Moreover, the I-V curves are strongly temperature-dependent and vanish above T* ~ 20 K (**Figs. 4c-4d**), consistent with the current sensitivity of resistivity that is strong below T* but negligible above T* (**Fig. 3c**). Note that at higher frequencies, an additional sharp jump $V_A$ emerges in the *c*-axis I-V curves (**Fig. 4d**), most prominently at 97.6 Hz. The absence of $V_A$ at lower frequencies and its strong anisotropy argue against a static pinning origin and point to a frequency-driven instability of the collective transport mode.

A strong frequency (*f*) dependence and an unusually long relaxation time are also distinct features of this instability. As shown in **Figs. 5a-5b**, both the positions of $V_{NDR}$ and the overall I-V nonlinearity depend strongly on $f$, shifting to higher V with increasing $f$; e.g., the $a$-axis I-V curve at $f$ = 97.6 Hz is vastly different from those at $f \leq$ 24.4 Hz **(Fig.5a)**. The extracted $V_{NDR}(f)$ trends (**Figs. 5c-5d**) reveal a well-defined knee frequency $f_c$ ~ 1-10 Hz, separating a low-$f$ regime where $V_{DP}$ and $V_{NDR}$ rapidly increase with increasing $f$ from a high-$f$ regime where depinning and sliding only weakly respond to further changes in $f$. The inferred relaxation time $\tau \sim (2\pi f_c)^{-1}$ is on the order of $10^{-3}$ s, several orders of magnitude longer than typical electronic scattering times (~ $10^{-15}$ - $10^{-12}$ s [1]). This anomalously slow dynamics provides compelling evidence that the nonlinear transport arises from collective motion of a strongly pinned density wave.

More insight is provided by the differential resistance, dV/dI, at an AC modulation of 1 mA in **Figs. 5e-5f**. At 5 K and 0.3 $\leq f \leq$ 18 Hz, dV/dI exhibits pronounced low $f$-dependent peak-dip structures and regions of NDR, which vanish at 18 Hz, once again signaling slow relaxation and non-adiabatic response of the density-wave instability (**Fig. 5e**). These features are completely absent above T* (**Fig. 5f**), where dV/dI becomes smooth, consistent with Ohmic transport.

The density-wave instability uncovered here differs fundamentally from conventional Peierls-type CDWs. Most notably, the simultaneous occurrence of pronounced $V_{DP}$, $V_{NDR}$ and slow collective dynamics in a magnetic oxide is unprecedented (**Fig. 4**). The electronically driven transition at $T_A$ is insensitive to magnetic fields (**Figs. 2a, 2b, 2d**) and is accompanied by only a small entropy removal (**Fig. 2d**), indicating a partial electronic reconstruction rather than spin-driven ordering. The absence of Curie–Weiss behavior in $\chi$(T) (**Figs. 2a-2c**), along with clear anomalies in lattice parameters at $T_A$ (**Figs. 1e-1g,** SFig.1), supports an electronic instability that is coupled to the lattice. A defining feature of this density-wave instability is the wide separation between its thermodynamic onset at $T_A$ and the emergence of collective dynamics below T* (**Fig. 1a**), in contrast to classical CDWs where depinning and sliding occur immediately below the ordering temperature. Our results highlight the unique role of trimers in hosting electronic instabilities that remain hidden at high temperatures where thermally activated quasiparticles dominate transport and become dynamically manifest only below T* where quasiparticles freeze out, thus unveiling nonlinearity in the collective mode (**Fig.1a**).

The mechanism driving this hidden density-wave instability remains to be established. However, the electronic configuration of Ru in $Ba_4Ru_3O_{10}$ may involve significant ligand-hole character ($d^5\underline{L}$) [44-46], which can support the coexistence of antiferromagnetism and itinerant charge degrees of freedom. Such a scenario is consistent with the emergence of collective charge dynamics within an antiferromagnetic background observed here.

**Acknowledgements**

This work is supported by U.S. National Science Foundation via Grant No. DMR 2204811. RN was supported by the U.S. National Science Foundation via Grant No DMR-2516302. The synchrotron X-ray work was supported by the U.S. Department of Energy, Office of Science, Basic Energy Sciences, Materials Sciences and Engineering Division. Work at the Center for High-Energy X-ray Sciences (CHEXS), which was supported by U.S. NSF under award DMR-2342336.

**Figure Captions**

**Fig. 1. Structural properties of $Ba_4Ru_3O_{10}$. a,** Schematic phase diagram illustrating the AFM + density wave electronic reconstruction (modulated red dots) below $T_A$. Transport is dominated by quasiparticles (moving purple dots) above T* but by the density-wave instability (moving modulated red dots) as quasiparticles freeze out below T*. **b,** Trimer of three face-sharing $RuO_6$ octahedra. **c-d,** Crystal structure in the *bc* plane (c) and in the *ac* plane (d). **e-g**, Temperature dependence of unit cell V (e), bond distance Ru-Ru (f) and bond angle Ru-O-Ru (g).

**Fig. 2. Magnetic susceptibility χ(T), Isothermal magnetization M(H), and heat capacity C(T) of $Ba_4Ru_3O_{10}$. a,** Temperature dependence of χ(T) along the *a*, *b*, and *c* axes at $\mu_oH$ = 0.5 T. **b**, Temperature dependence of $\chi_a$(T) at $\mu_oH$ = 0.1, 0.5, and 5 T. $T_A$ remains unchanged. **c,** *a*-axis $M_a$(H) at T = 1.8 and 150 K. Note $M_a$(1.8 K) < $M_a$(150 K). **d,** C(T) at $\mu_oH$ = 0 and 14 T in the vicinity of $T_A$ with an entropy change ΔS ≈ 0.4 J/mole K. $T_A$ is field independent. **e,** C(T) at $\mu_0H$ = 0 and 14 T for a wider range 1.8 ≤ T ≤ 200K. **f,** C/T vs $T^2$ for 1.8 ≤ T ≤ 10 K showing the absence of the $T^3$-term anticipated for a conventional AFM ground state.

**Fig. 3. Current-dependent transport and suppression of nonlinear transport by Ir substitution in $Ba_4(Ru_{1-x}Ir_x)_3O_{10}$. a-b,** Temperature dependence of $\chi_a$(T) at $\mu_oH$ = 0.1 T(a) and the *a*-axis resistivity $\rho_a$(T) at I = 0.1 mA (b) for x = 0 and x = 0.03. **c,** $\rho_a$(T) for x = 0 with increasing I revealing strong current dependence below T* ~ 20 K (shaded area). **Inset:** ρ(T)/ρ(300K) shows that the anisotropy between $\rho_a$(T) and $\rho_c$(T) is visible near the $T_A$ regime but vanishes below T*. **d,** Corresponding $\rho_a$(T) for x = 0.03 remaining nearly current independent. **e,** Difference in resistivity $\Delta\rho_a$ = $\rho_a$(0.1mA) – $\rho_a$(5mA) highlighting the strong current sensitivity in x = 0 below T* and its suppression upon Ir substitution. **f,** I-V curves at 10 K for x = 0 and x = 0.03, respectively, showing depinning $V_{DP}$ and $V_{NDR}$ only in x = 0. The contrasting behavior between x = 0 and x = 0.03 demonstrates the intrinsic nature of the nonlinearity in x = 0.

**Fig. 4. Anisotropic collective transport and electric-field scaling in $Ba_4Ru_3O_{10}$. a-b,** I-V characteristics at 10 K and 24.2 Hz for the *a* and *c* axes. **b,** Corresponding J-E curves derived from the data in (a), demonstrating the persisting, large anisotropy after eliminating geometric effects. **c,** Temperature evolution of the I-V curves along the *a* axis showing the suppression of nonlinear

features with increasing T. **d,** High-*f* *c*-axis I-V curves revealing an additional sharp anomaly $V_A$ at 97.6 Hz.

**Figure 5. Frequency-dependent nonlinear transport, knee frequency and differential resistance of the density-wave state in $Ba_4Ru_3O_{10}$. a-b,** I-V curves along the *a* and *c* axes at 10 K for different *f*, showing systematic shifts of $V_{NDR}$ to higher V with increasing *f*. **c-d,** Extracted $V_{NDR}$ as a function of *f* for the *a* and *c* axes, respectively, revealing a knee $f_c$ (1-10 Hz) reflects an extraordinarily long relaxation time of the density wave. **e,** Differential resistance dV/dI along the *a* axis at 5 K for different *f*, displaying pronounced *f*-dependent peak-dip structures and regions of NDR associated with collective depinning and sliding. **f,** Temperature dependence of the *a*-axis dV/dI at 1.5 Hz, showing that the nonlinear features present at 5 and 10 K but vanishing completely by 20 K. Note that dV/dI was measured using a small AC modulation superimposed on a DC bias current.

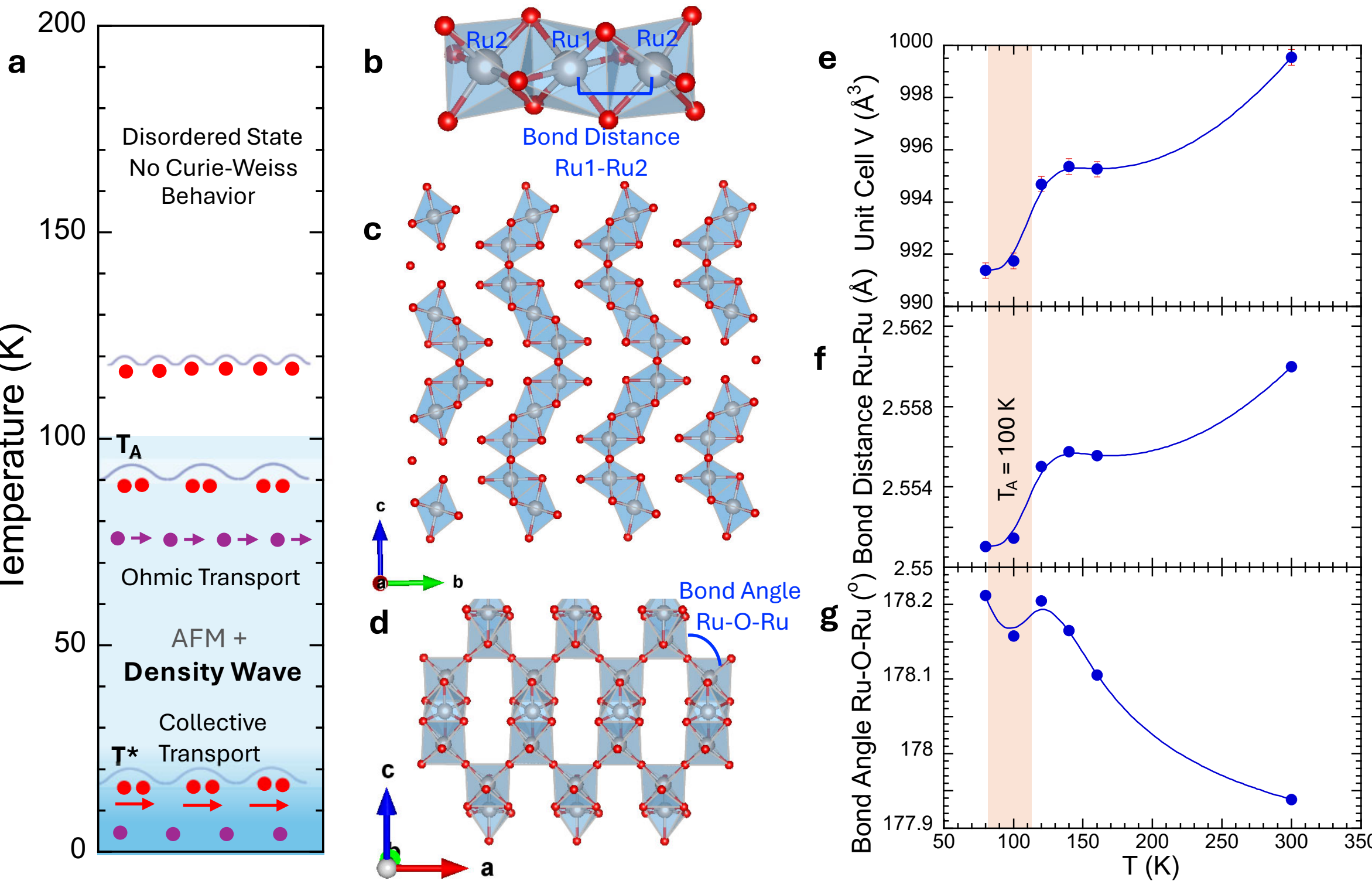

Figure 1

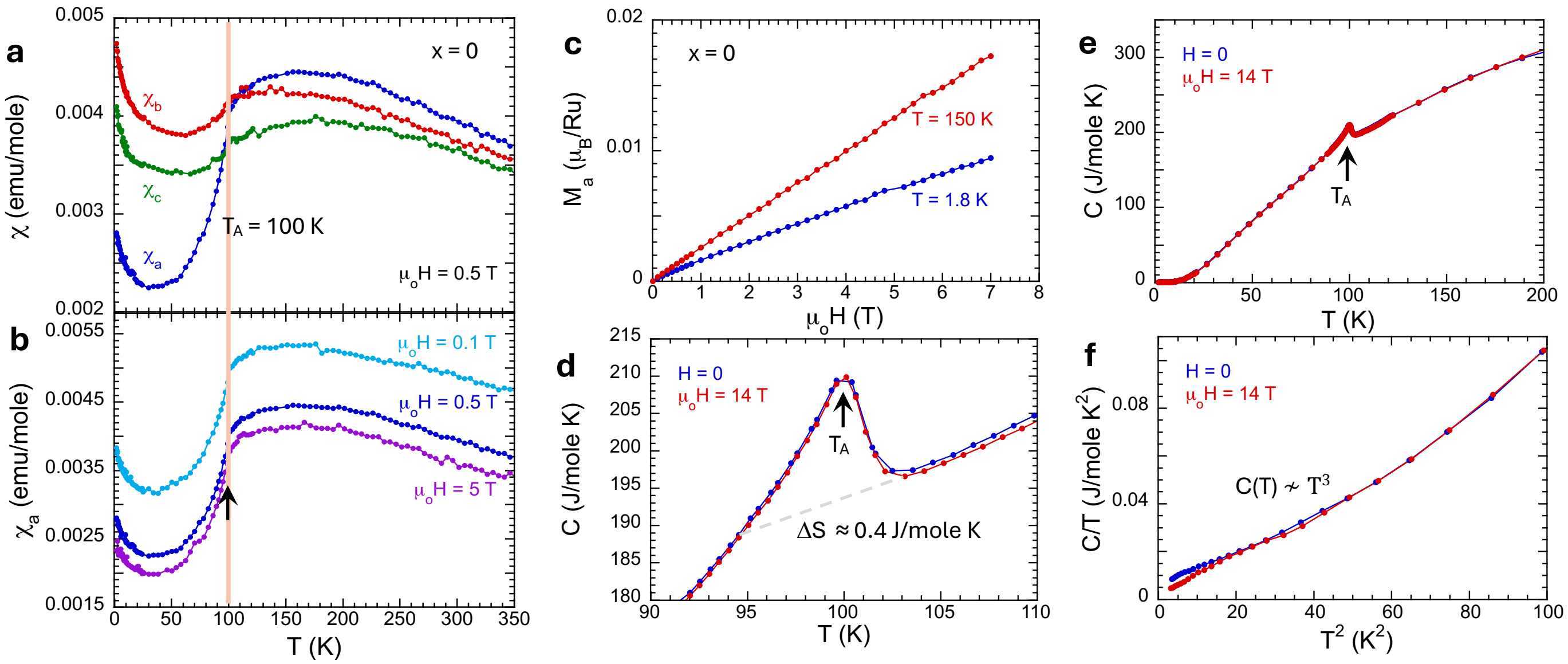

Figure 2

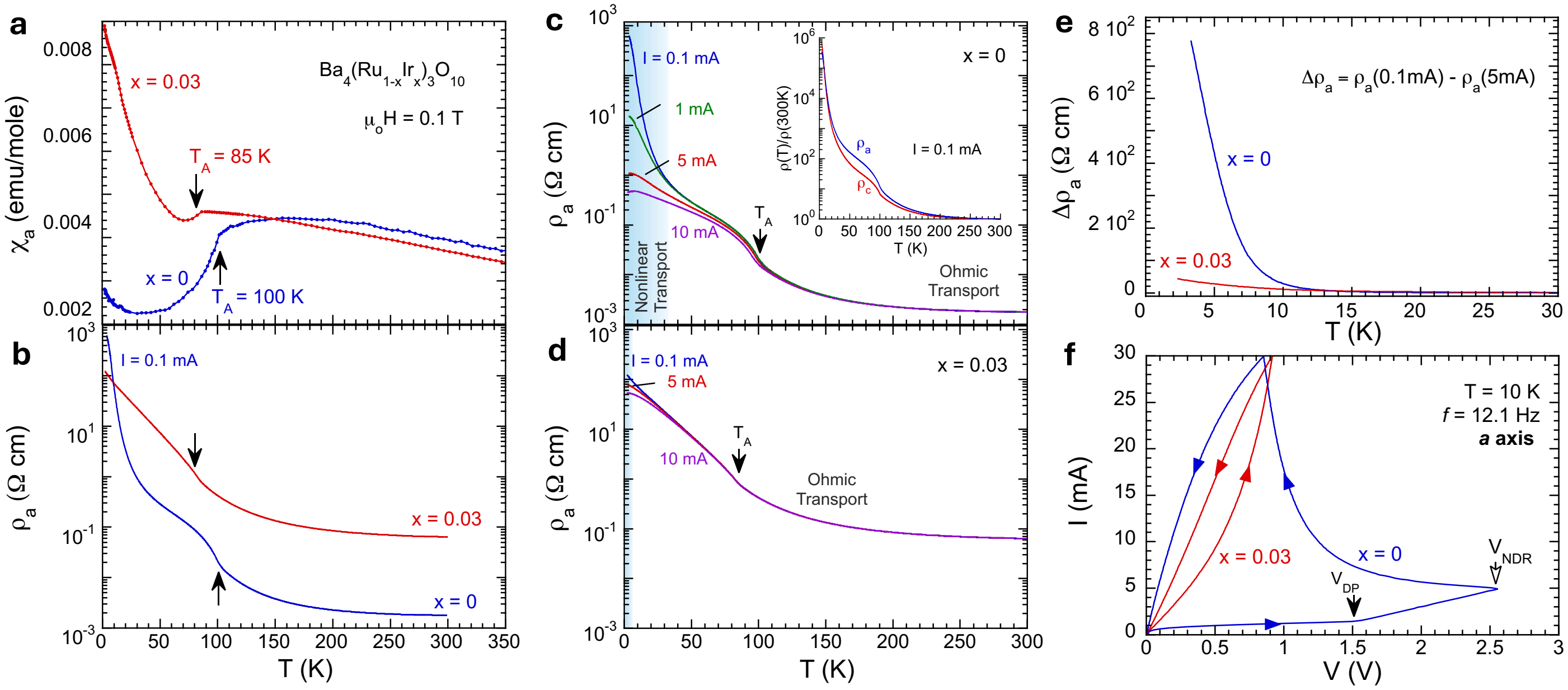

Figure 3

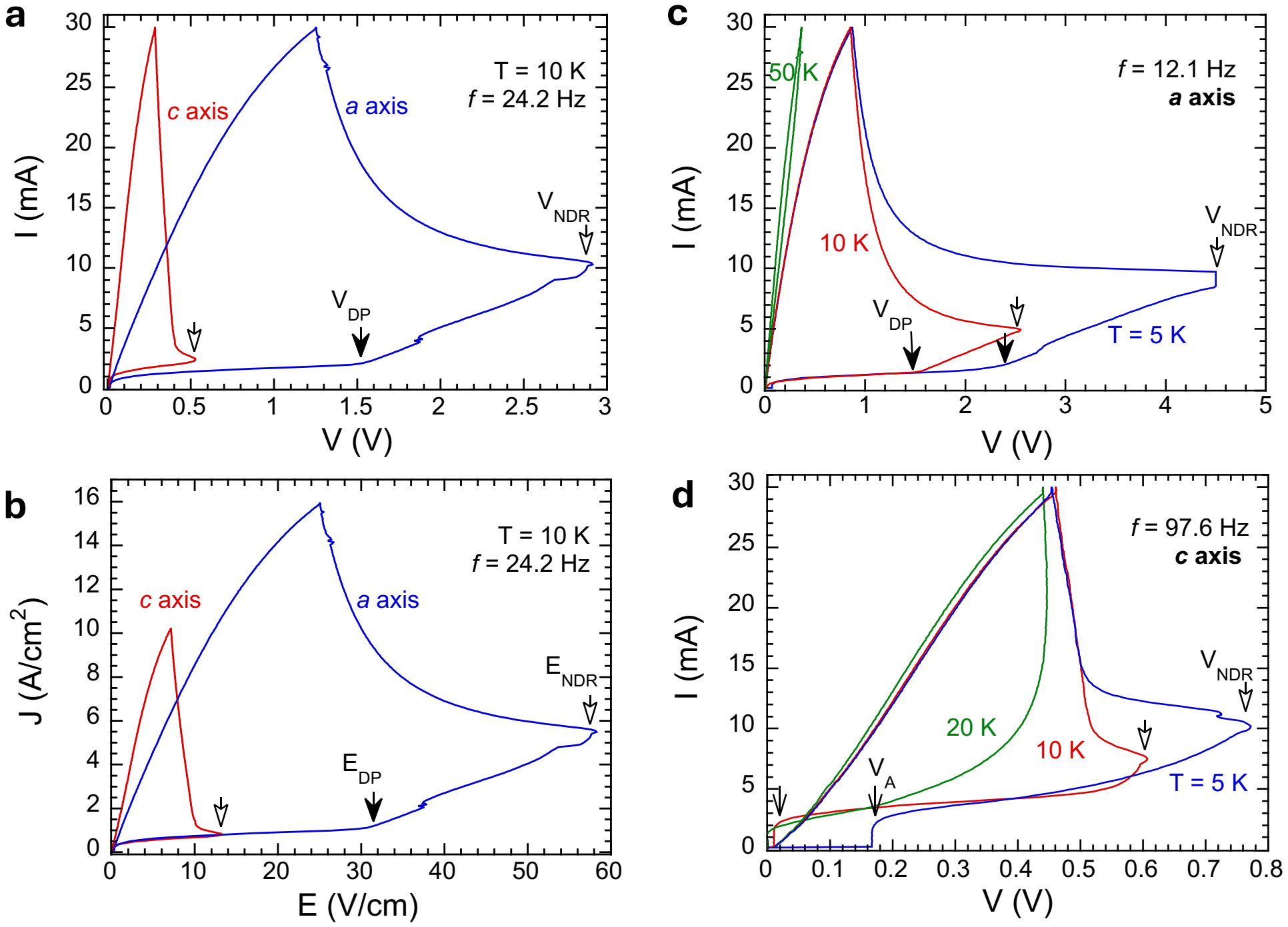

Figure 4

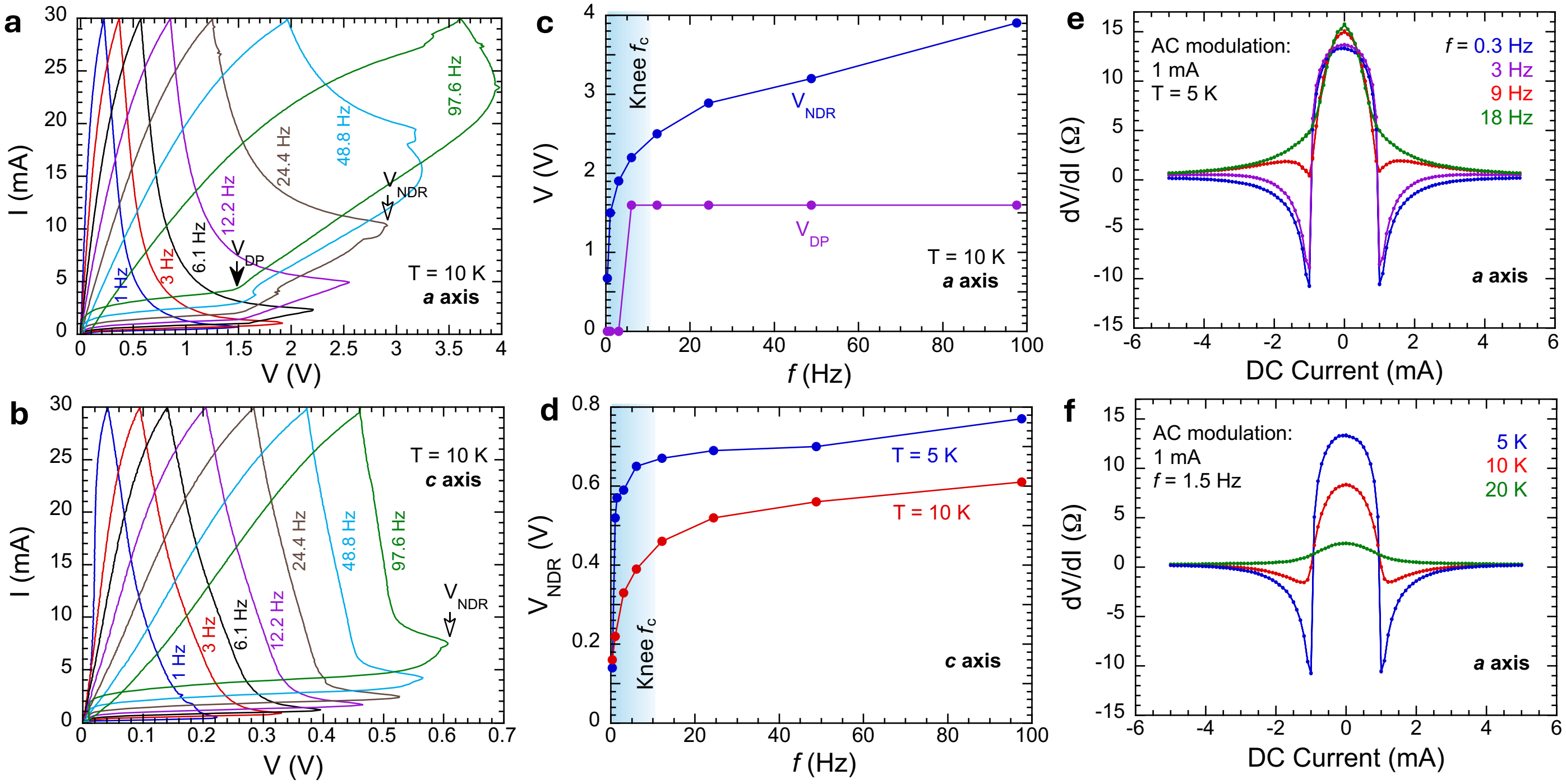

Figure 5

Supplemental Material

# Hidden Density-Wave Instability in the Trimer Ruthenate $Ba_4Ru_3O_{10}$

Gang Cao[1*], Hengdi Zhao[1,2], Adrienne Bond[1], Tristan R. Cao[1], Gabriel Schebel[1], Arabella Quane[1], Yifei Ni[1], Yu Zhang[1], Logan Wall[1], Rahul Nandkishore[1], Pedro Schlottmann[3], Stephan Rosenkranz[2], and Feng Ye[4]

[1]Department of Physics, University of Colorado at Boulder, Boulder, CO 80309, USA

[2]Materials Science Division, Argonne National Laboratory, Lemont, IL 60439, USA

[3]Department of Physics, Florida State University, Tallahassee, FL 32306, USA

[4]Oak Ridge National Laboratory, Oak Ridge, TN 37830, USA

## I. Experimental

Single crystals of $Ba_4Ru_3O_{10}$ were grown using a flux method. Measurements of crystal structures were performed using a Bruker Quest ECO single-crystal diffractometer with an Oxford Cryosystem providing sample temperature environments ranging from 80 K to 400 K. Chemical analyses of the samples were performed using a combination of a Hitachi MT3030 Plus Scanning Electron Microscope and an Oxford Energy Dispersive X-Ray Spectroscopy (EDX). The measurements of the electrical resistivity, I-V characteristics and heat capacity were carried out using a Quantum Design (QD) Dynacool PPMS system having a 14-Tesla magnet and a set of external Keithley meters that provides current source and measures voltage with a high precision. Note that the I-V curves are current-driven. The contact resistance was measured to be on the order of 1 Ω, and these measurements were made using standard four-probe configurations, with current and voltage leads separated to eliminate contact contributions to voltage measurements.

## II. Additional Data

*1. Lattice parameters of $Ba_4Ru_3O_{10}$*

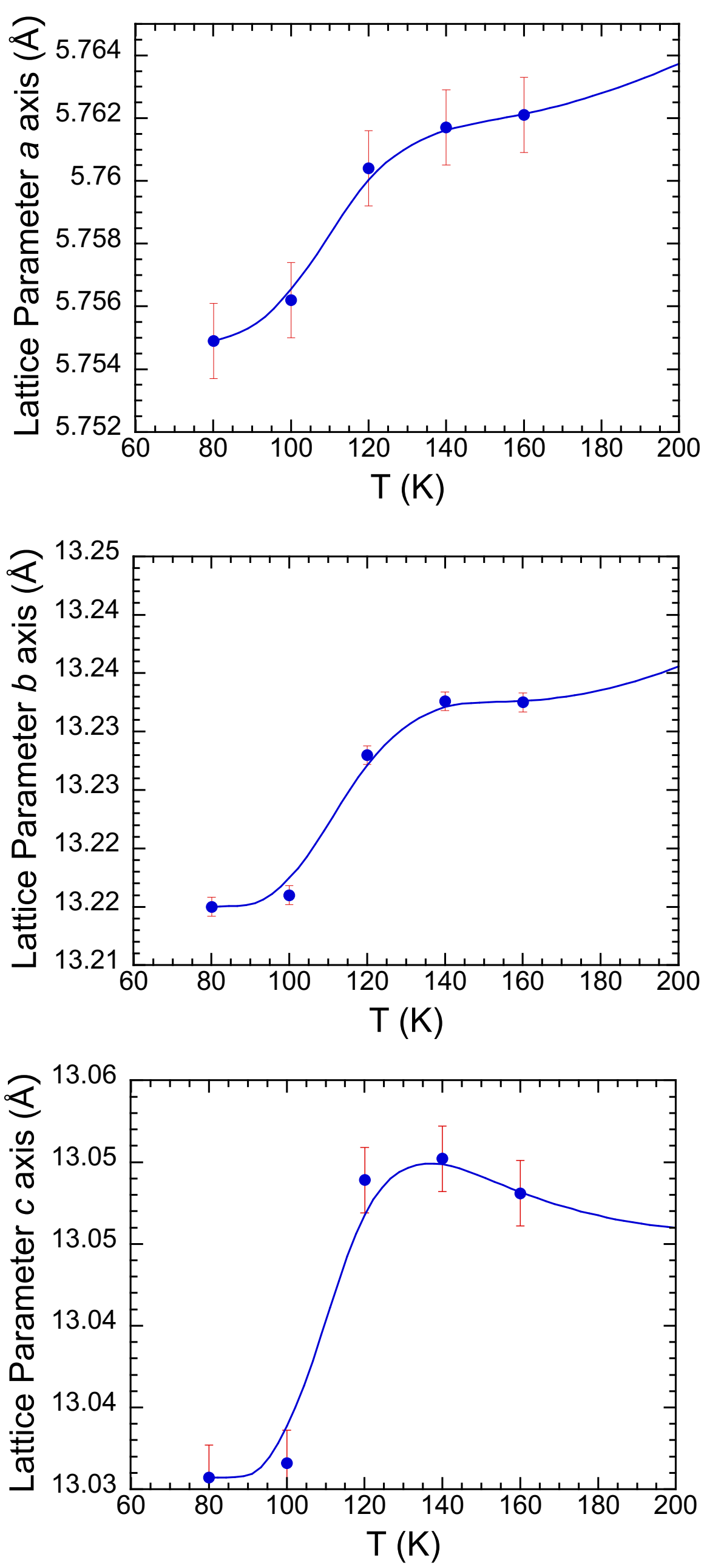

***SFig. 1. Structural properties of $Ba_4Ru_3O_{10}$:*** *The lattice parameters a, b and c axes as a function of temperature showing anomalies near 100 K.*

*2. Isothermal Magnetization of $Ba_4(Ru_{1-x}Ir_x)_3O_{10}$ with x =0.03*

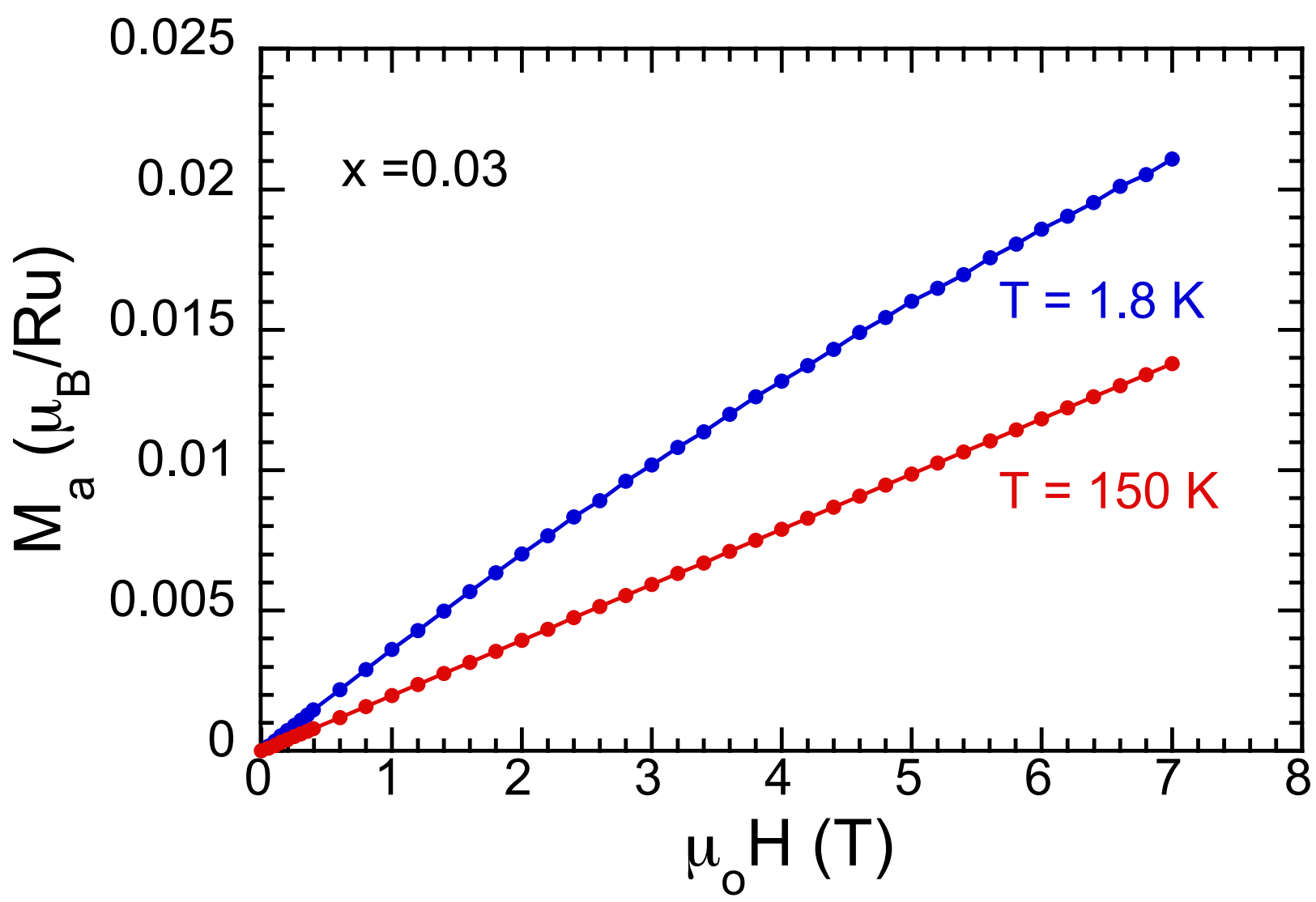

***SFig.2. a-axis Isothermal magnetization M(H) at 1.8 and 150 K for x = 0.03.*** *Note that M(H) behaves normally, sharply contrasting that for x = 0 where Ma(1.8 K) < Ma(150 K) (Fig. 2c in the main text).*

**III. Additional Data and Arguments on Joule Heating**

We have directly measured the temperature rise of the sample under applied currents using a calibrated Cernox thermometer (**SFig. 3**). These measurements show that the temperature increase remains very small: $\Delta T < 3$ K at 10 K even at I = 10 mA and is negligible (<1 K) at I = 2mA, where pronounced nonlinear transport is already observed (**Fig. 4a**).

Such small temperature variations are insufficient to account for the observed behavior. In particular, the resistivity $\rho(T)$ below 30 K evolves smoothly and monotonically (**Fig. 3b**), so a temperature shift of only a few Kelvin cannot generate the sharp nonlinearities and negative differential resistance observed in the I-V characteristics (**Figs. 3-5**).

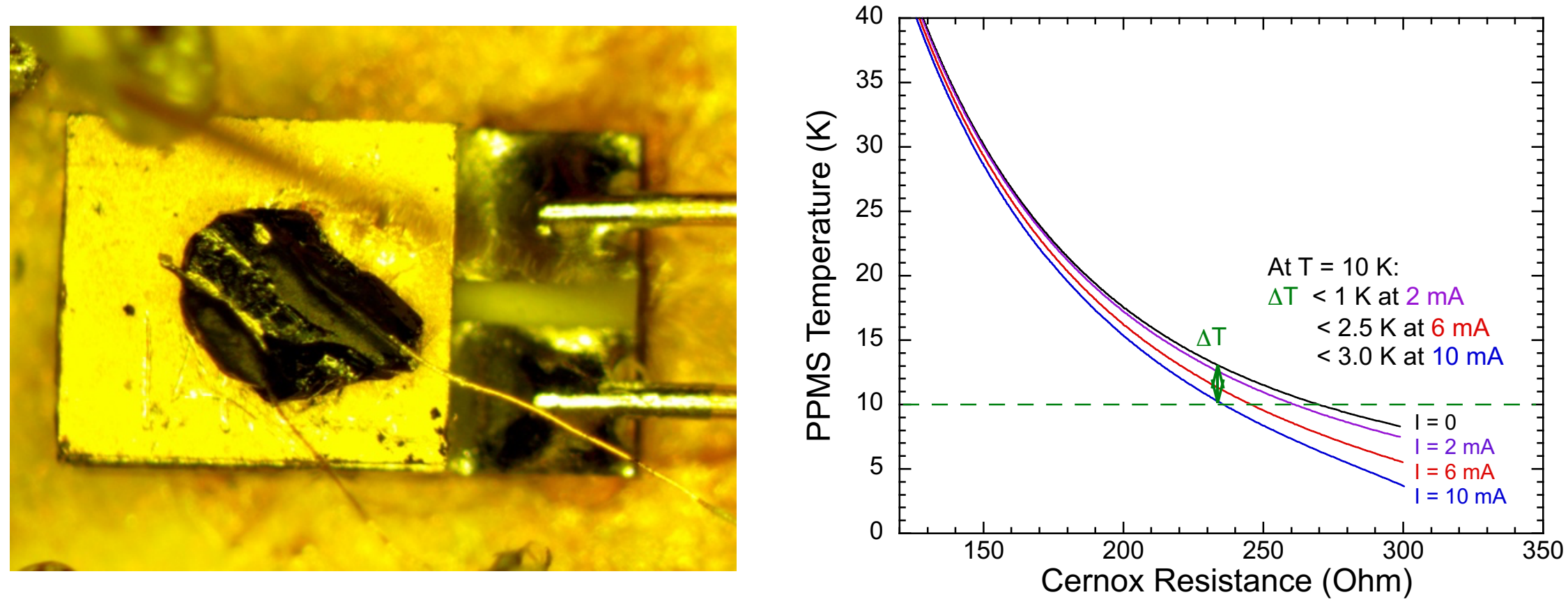

***SFig. 3. Direct measurements of current-induced temperature changes.*** *Left: Optical image of a single-crystal sample (black) mounted on a Cernox thermometer (gold). Right: Temperature as a function of Cernox resistance for different applied currents I = 0 (calibration), 2 mA, 6 mA, and 10 mA. The inferred temperature rise remains small, with DT < 3 K at 10 K even for I = 10 mA and negligible (<1 K) at I = 2 mA. These results demonstrate that Joule heating is minimal in the current range where strong nonlinear transport is observed.*

Equally importantly, several independent observations are incompatible with a heating scenario:

1. ***Absence of nonlinear transport in insulting*** $Ba_4(Ru_{0.97}Ir_{0.03})_3O_{10}$**:**

   The 3% Ir-substituted sample remains insulating (**Fig. 3d**) but shows no nonlinear transport (**Fig.3f**), demonstrating that the effect is highly sensitive to the electronic state and not to measurement conditions.

2. ***Strong anisotropy:***

   The nonlinear I-V characteristics are highly anisotropic (**Fig. 4**), whereas Joule heating would produce an isotropic response.

3. ***Low-current onset:***

   Significant nonlinearity already appears at currents as low as 2 mA along the *c* axis (**Fig. 4a**), where the measured temperature rise is negligible.

4. ***Sharp thresholds and NDR***

Negative differential resistance emerges at well-defined current thresholds rather than gradually as expected for thermal effects (**Figs. 4-5**).

5. ***Dynamic signatures and strong frequency dependence:***

   The I-V curves and differential resistance measurements (**Fig. 5**) exhibit strong current- and frequency-dependent behavior, indicative of slow collective dynamics that cannot be explained by heating.

All in all, the direct temperature measurements and the transport characteristics rule out Joule heating as the origin of the observed nonlinear behavior. Instead, they support the interpretation as arising from an intrinsic, strongly pinned collective electronic mode.

## IV. Additional X-ray Data from 80 K to 20 K

**Table S1 Crystal data and structure refinement for $Ba_4Ru_3O_{10}$ at 80K.**

| | |
|---|---|
| Empirical formula | $Ba_4O_{10}Ru_3$ |
| Formula weight | 1012.57 |
| Temperature/K | 80 |
| Crystal system | orthorhombic |
| Space group | Cmce |
| a/Å | 5.7612(8) |
| b/Å | 13.233(2) |
| c/Å | 13.039(2) |
| α/° | 90 |
| β/° | 90 |
| γ/° | 90 |
| Volume/Å$^3$ | 994.1(3) |
| Z | 4 |
| $\rho_{calc}$g/cm$^3$ | 6.766 |
| μ/mm$^{-1}$ | 6.698 |
| F(000) | 1744.0 |
| Radiation | Synchrotron X-ray (λ = 0.24797) |
| 2Θ range for data collection/° | 2.148 to 28.708 |
| Index ranges | $-11 \leq h \leq 11$, $-26 \leq k \leq 26$, $-25 \leq l \leq 25$ |
| Reflections collected | 41834 |
| Independent reflections | 2220 [$R_{int}$ = 0.0536, $R_{sigma}$ = 0.0177] |
| Data/restraints/parameters | 2220/0/50 |

| Goodness-of-fit on $F^2$ | 1.196 |
|---|---|
| Final R indexes [I>=2σ (I)] | $R_1$ = 0.0258, $wR_2$ = 0.0655 |
| Final R indexes [all data] | $R_1$ = 0.0267, $wR_2$ = 0.0696 |
| Largest diff. peak/hole / e $Å^{-3}$ | 5.48/-4.75 |

**Table S1 (Cont'd) Extended Fractional Atomic Coordinates (×$10^4$) and Equivalent Isotropic Displacement Parameters ($Å^2$×$10^3$) for $Ba_4Ru_3O_{10}$ at 80K. $U_{eq}$ is defined as 1/3 of the trace of the orthogonalised $U_{IJ}$ tensor.**

| **Atom** | ***x*** | ***y*** | ***z*** | **U(eq)** |
|---|---|---|---|---|
| Ba1 | 5000 | 351.8(2) | 1388.2(2) | 2.73(4) |
| Ba2 | 0 | 2394.7(2) | 1115.8(2) | 3.70(4) |
| Ru1 | 0 | 0 | 0 | 2.67(5) |
| Ru4 | 0 | -1247.8(2) | 1494.0(2) | 2.45(5) |
| O1 | -2500 | -1225.6(14) | 2500 | 5.5(2) |
| O2 | 0 | 343.4(14) | 1523.2(14) | 4.7(2) |
| O3 | 0 | -2698.1(15) | 1470.2(16) | 6.4(3) |
| O4 | -2294(2) | 1098.6(10) | -347.3(9) | 4.45(15) |

**Table S2 Crystal data and structure refinement for $Ba_4Ru_3O_{10}$ at 50K.**

| Empirical formula | $Ba_4O_{10}Ru_3$ |
|---|---|
| Formula weight | 1012.57 |
| Temperature/K | 50 |
| Crystal system | orthorhombic |
| Space group | Cmce |
| a/Å | 5.7590(9) |
| b/Å | 13.232(2) |
| c/Å | 13.037(2) |
| α/° | 90 |
| β/° | 90 |
| γ/° | 90 |
| Volume/$Å^3$ | 993.4(3) |
| Z | 4 |
| $\rho_{calc}$g/cm$^3$ | 6.770 |
| μ/mm$^{-1}$ | 6.702 |
| F(000) | 1744.0 |
| Radiation | Synchrotron X-ray (λ = 0.24797) |
| 2Θ range for data collection/° | 2.148 to 28.714 |
| Index ranges | $-11 \leq h \leq 11$, $-26 \leq k \leq 26$, $-25 \leq l \leq 25$ |
| Reflections collected | 64121 |
| Independent reflections | 2221 [$R_{int}$ = 0.0563, $R_{sigma}$ = 0.0146] |
| Data/restraints/parameters | 2221/0/50 |

| Goodness-of-fit on $F^2$ | 1.162 |
|---|---|
| Final R indexes [I>=2σ (I)] | $R_1$ = 0.0258, $wR_2$ = 0.0682 |
| Final R indexes [all data] | $R_1$ = 0.0261, $wR_2$ = 0.0749 |
| Largest diff. peak/hole / e $Å^{-3}$ | 5.29/-5.13 |

**Table S2 (Cont'd) Extended Fractional Atomic Coordinates (×10$^4$) and Equivalent Isotropic Displacement Parameters (Å$^2$×10$^3$) for $Ba_4Ru_3O_{10}$ at 50K. $U_{eq}$ is defined as 1/3 of the trace of the orthogonalised $U_{IJ}$ tensor.**

| Atom | *x* | *y* | *z* | U(eq) |
|---|---|---|---|---|
| Ba1 | 5000 | 351.9(2) | 1388.2(2) | 2.74(5) |
| Ba2 | 0 | 2394.3(2) | 1115.8(2) | 3.64(5) |
| Ru1 | 0 | 0 | 0 | 2.76(5) |
| Ru4 | 0 | -1247.6(2) | 1493.7(2) | 2.59(5) |
| O1 | -2500 | -1225.9(17) | 2500 | 5.2(3) |
| O2 | 0 | 344.2(17) | 1523.6(17) | 4.7(3) |
| O3 | 0 | -2698.9(18) | 1471.8(19) | 5.7(3) |
| O4 | -2298(3) | 1099.9(11) | -347.6(12) | 4.66(19) |

**Table S3 Crystal data and structure refinement for $Ba_4Ru_3O_{10}$ at 20K.**

| Empirical formula | $Ba_4O_{10}Ru_3$ |
|---|---|
| Formula weight | 1012.57 |
| Temperature/K | 20 |
| Crystal system | orthorhombic |
| Space group | Cmce |
| a/Å | 5.7594(7) |
| b/Å | 13.2284(16) |
| c/Å | 13.0343(16) |
| α/° | 90 |
| β/° | 90 |
| γ/° | 90 |
| Volume/$Å^3$ | 993.1(2) |
| Z | 4 |
| $\rho_{calc}$g/cm$^3$ | 6.773 |
| μ/mm$^{-1}$ | 6.704 |
| F(000) | 1744.0 |
| Radiation | Synchrotron X-ray (λ = 0.24797) |
| 2Θ range for data collection/° | 2.148 to 28.718 |
| Index ranges | -11 ≤ h ≤ 11, -26 ≤ k ≤ 26, -25 ≤ l ≤ 25 |
| Reflections collected | 24659 |
| Independent reflections | 2214 [$R_{int}$ = 0.0539, $R_{sigma}$ = 0.0231] |
| Data/restraints/parameters | 2214/0/50 |

| Goodness-of-fit on $F^2$ | 1.125 |
|---|---|
| Final R indexes [I>=2σ (I)] | $R_1$ = 0.0360, $wR_2$ = 0.0974 |
| Final R indexes [all data] | $R_1$ = 0.0363, $wR_2$ = 0.1031 |
| Largest diff. peak/hole / e $Å^{-3}$ | 6.67/-5.12 |

**Table S3 (Cont'd) Extended Fractional Atomic Coordinates (×$10^4$) and Equivalent Isotropic Displacement Parameters ($Å^2$×$10^3$) for $Ba_4Ru_3O_{10}$ at 20K. $U_{eq}$ is defined as 1/3 of the trace of the orthogonalised $U_{IJ}$ tensor.**

| **Atom** | ***x*** | ***y*** | ***z*** | **U(eq)** |
|---|---|---|---|---|
| Ba1 | 5000 | 351.8(2) | 1388.5(2) | 3.40(7) |
| Ba2 | 0 | 2394.9(2) | 1115.9(2) | 4.43(7) |
| Ru1 | 0 | 0 | 0 | 3.30(7) |
| Ru4 | 0 | -1247.7(2) | 1493.7(2) | 3.07(7) |
| O1 | -2500 | -1228(2) | 2500 | 6.2(3) |
| O2 | 0 | 341(2) | 1527(2) | 5.6(3) |
| O3 | 0 | -2704(2) | 1473(2) | 7.1(4) |
| O4 | -2296(3) | 1097.7(13) | -347.4(13) | 4.8(2) |